\documentclass[aps,pra,preprint,groupedaddress,showkeys]{revtex4-1}
\usepackage{graphicx}
\usepackage{bm}
\usepackage{amsmath}
\usepackage{subfigure}

\newcommand{\ket}[1]{\left| #1 \right\rangle}

\newcommand{\bra}[1]{\left\langle #1 \right|}

\begin{document}

\title{Characterizing a configuration interaction excited state using natural transition geminals}

\author{J. P. Coe}
\author{M. J. Paterson}%
\affiliation{ 
Institute of Chemical Sciences, School of Engineering and Physical Sciences, Heriot-Watt University, Edinburgh, EH14 4AS, UK.
}%

\begin{abstract}
We introduce natural transition geminals as a means to qualitatively understand a transition where double excitations are important. The first two $A_{1}$ singlet states of the CH cation are used as an initial example.  We calculate these states with configuration interaction singles (CIS) and state-averaged Monte Carlo configuration interaction (SA-MCCI).  For each method we compare the important natural transition geminals with the dominant natural transition orbitals. We then compare SA-MCCI and full configuration interaction (FCI) with regards to the natural transition geminals using the beryllium atom.  We compare using the natural transition geminals with analyzing the important configurations in the CI expansion to give the dominant transition for the beryllium atom and the carbon dimer.  Finally we calculate the natural transition geminals for two electronic excitations of formamide.
\end{abstract}


\maketitle

\section {Introduction}

Using the important transitions from the ground-state to characterize an excited state offers a qualitative interpretation of the excitation.  Natural transition orbitals \cite{Martin2003} have been shown to give a compact description of the important orbitals involved in an electronic excitation when modelled using single substitutions.  For configuration interaction wavefunctions, however, double excitations may be dominant
and these will not be detected when using natural transition orbitals.  We therefore put forward the concept of natural transition geminals which will categorize the important two electron transitions.  As the one-particle transition density matrix may be calculated from the two-particle transition density matrix then the natural geminals will also contain information regarding single-particle transitions.  Therefore they offer the possibility of a more general approach for characterizing an electronic transition despite their added complexity and the difficulty in visualizing a function of six coordinates.  

Geminals, or antisymmetric two-electron functions, have been used in quantum chemistry methods since the 1950s and have a number of appealing features, see Ref.~\cite{SURJAN} for a review.  However these approaches perhaps did not become popular due to insufficient recovery of the correlation energy given their complexity.  More recently, there have been promising improvements on these methods \cite{Rassolov2002,Rassolov2004} and geminals have also been combined with explicitly correlated basis (F12) methods \cite{Valeev06}.  The natural geminals, defined as the two-electron eigenfunctions of the second-order
reduced density matrix, were used to investigate the beryllium atom in Ref.~\cite{Fogel65}.  However this paper represents the first application of natural transition geminals to our knowledge.

We test the approach of natural transition geminals on Monte Carlo configuration interaction (MCCI) \cite{mcciGreer98,mccicodeGreer} wavefunctions.  MCCI allows a wavefunction, which can reproduce much of the full configuration interaction (FCI) result, to be built up using usually only a small fraction of the configurations that would be necessary for FCI.  MCCI is an iterative scheme where the configuration space is stochastically enlarged and configurations that have an absolute coefficient less than a certain value ($c_{\text{min}}$) in the MCCI wavefunction are ultimately removed.  The method has been employed to successfully find compact wavefunctions for excitations of atoms and small molecules \cite{GreerMcciSpectra}, multipole moments, ionization energies and electron affinities \cite{MCCIdipoles}, and the potential curves of ground states \cite{MCCIpotentials}. We have recently used state-averaged MCCI \cite{MCCIstateAveraging} (SA-MCCI) to model excited potential curves with conical intersections or avoided crossings and the vertical excitations of small organic molecules.  In this paper we consider the ground and first singlet excited state of $A_{1}$ symmetry for the CH cation in the cc-pVDZ basis as an example to compare natural transition orbitals with natural transition geminals. We also contrast configuration interaction limited to single substitutions (CIS) and SA-MCCI on this system.  We then compare  the FCI natural transition geminals with those of SA-MCCI for the beryllium atom in the cc-pVDZ basis.  We compare using the dominant configurations in the CI expansion with the natural transition geminals as a means to infer the main transition for an excitation of the beryllium atom and the carbon dimer.  Finally we calculate the natural transition geminals for two excitations of $A'$ symmetry for formamide when using SA-MCCI and the def1-TZVP basis.

\section{Methods}

\subsection{MCCI}

In this work we use MCCI \cite{mcciGreer98,mccicodeGreer} with a form of state averaging (SA-MCCI) and employ Slater determinants (SDs).  We briefly describe the iterative scheme which begins with a Hartree-Fock SD.  SA-MCCI augments a set of SDs randomly with single and double substitutions which maintain the symmetry of the wavefunction. The Hamiltonian matrix is then constructed and diagonalized to give the first $s$ states.  We then create state-averaged coefficients using $c_{i}=\sum_{j=1}^{s}|c_{i,j}|$ and SDs that have just been included are removed if $|c_{i}|< c_{\text{min}}$   \cite{MCCIstateAveraging}.  This approach is repeated and every ten iterations all SDs become candidates for removal.

\subsection{Natural transition orbitals}

The natural transition orbitals \cite{Martin2003} are found by performing a singular value decomposition on the single-particle transition density matrix which, in spatial-spin co-ordinates, is 

\begin{equation}
T (\vec{x}_{A},\vec{x}_{B})=N\int \Psi_{2}^{*}(\vec{x}_{A},\vec{x}_{2},\cdots,\vec{x}_{N}) \Psi_{1}(\vec{x}_{B},\vec{x}_{2},\cdots,\vec{x}_{N})d\vec{x}_{2}\cdots\vec{x}_{N}.
\end{equation}

For two $N$ electron wavefunctions constructed from Slater determinants and the same set of molecular orbitals ($\phi_{1}$ to $\phi_{M}$) we can write the single-particle transition density matrix in terms of the spin orbitals ($\chi_{1}$ to $\chi_{2M}$):
\begin{equation}
T (\vec{x}_{A},\vec{x}_{B})=\sum_{i=1}^{2M}\sum_{j=1}^{2M}\chi_{i}^{*}(\vec{x}_{A})T_{ij}\chi_{j}(\vec{x}_B).
\end{equation}

The Slater determinants comprising the wavefunctions must have no more than one difference to contribute to $\bm{T}$ so we have 
\begin{equation}
T_{ij}=\bra{\Psi_{2}} c^{\dagger}_{i}c_{j} \ket{\Psi_{1}}
\end{equation}
where $i$ and $j$ range over all spin orbitals.  If we consider a ground state where all spatial orbitals are doubly occupied and a triplet state formed from a single spatial orbital replacement then the contributions to $\bm{T}$ would cancel if we were to sum over spins.  Therefore we do not average over spins here. Empty rows and columns in $\bm{T}$ may be removed and the matrix relabeled followed by a singular value decomposition $\bm{T}=\bm{U D V^{\dagger}}$ to give the natural transition orbitals and their eigenvalues for $\Psi_{2}$ and $\Psi_{1}$.

We may build the single-particle transition density matrix for two MCCI wavefunctions $\Psi_{2}=\sum_{i} c_{i} SD_{i}$ and $\Psi_{1}=\sum_{j} d_{j} SD_{j}$  by noting that SDs $i$ and $j$ in maximum coincidence only contribute if there is one difference with orbitals $k$ and $l$ responsible
\begin{equation}
T_{kl}\rightarrow T_{kl}+e_{p}c_{i}^{*}d_{j}.
\end{equation}
or no differences
\begin{equation}
T_{mm}\rightarrow T_{mm}+e_{p}c_{i}^{*}d_{j}.
\end{equation}
Here $m$ runs over all spin orbitals in either SD and $e_{p}$ is the sign from placing the Slater determinants in maximum coincidence.

\subsection{Natural transition geminals}

We construct natural transition geminals by first considering the two-particle transition density matrix

\begin{eqnarray}
\nonumber T^{(2)} (\vec{x}_{A},\vec{x}_{A'},\vec{x}_{B},\vec{x}_{B'})=\\
\frac{N(N-1)}{2}\int \Psi_{2}^{*}(\vec{x}_{A},\vec{x}_{A'},\vec{x}_{3},\cdots,\vec{x}_{N}) \Psi_{1}(\vec{x}_{B},\vec{x}_{B'},\vec{x}_{3},\cdots,\vec{x}_{N})d\vec{x}_{3}\cdots\vec{x}_{N}
\end{eqnarray}

which can be written as
\begin{equation}
T^{(2)} (\vec{x}_{A},\vec{x}_{A'},\vec{x}_{B},\vec{x}_{B'})=\sum_{i=1}^{M_{G}}\sum_{j=1}^{M_{G}}G_{i}^{*}(\vec{x}_{A},\vec{x}_{A'})T^{(2)}_{ij}G_{j}(\vec{x}_B,\vec{x}_{B'})
\end{equation}

where the $G_{i}$ are the $\binom{2M}{2}=M_{G}$ unique geminals formed from pairs of spin orbitals.  For example

\begin{equation}
 G_{1}(\vec{x}_{1},\vec{x}_{2})=\ket{1,\bar{2}}.
\end{equation}

Here the ket at the right hand side is a normalized SD and the $1$ without a bar represents $\phi_{1}$ with an up spin while with a bar it would have a down spin.

The two particle transition matrix in terms of geminals can be constructed as
\begin{equation}
T^{(2)}_{ij}=\bra{\Psi_{2}} c^{\dagger}_{i_1} c^{\dagger}_{i_2}c_{j_1}c_{j_2} \ket{\Psi_{1}}.
\end{equation}

Here we label the spin orbitals with spin up using the molecular orbital number ($1$ to $M$) while the spin down molecular orbitals are labelled from $M+1$ to $2M$ using the molecular orbital number plus $M$.
We stipulate that $i_{1} < i_{2}$ and $j_{1} < j_{2}$ for our list of unique geminals which we map from two labels to one using

\begin{equation}
i(i_{1},i_{2})=2(i_{1}-1)M+i_{2}-i_{1}-\frac{(i_{1}-1)i_{1}}{2}.
\end{equation}
An example of the geminal labelling scheme for two MOs is given in table~\ref{Tbl:example}.

\begin{table}
  \caption{Example labelling for geminals when using two molecular orbitals.}
{\begin{tabular}{@{}cccc}\toprule
  Geminal & $i_{1}$ & $i_{2}$
         & $i$\\
\colrule
 $\ket{1,2}$ & 1 & 2 & 1\\
 $\ket{1,\bar{1}}$  & 1 & 3 & 2 \\
 $\ket{1,\bar{2}}$ & 1 & 4 & 3 \\
 $\ket{2,\bar{1}}$ & 2 & 3 & 4 \\
$\ket{2,\bar{2}}$& 2 & 4 & 5 \\
$\ket{\bar{1},\bar{2}}$ & 3 & 4 & 6 \\
   \botrule
  \end{tabular}}
\label{Tbl:example}
\end{table}

Similarly to the procedure for the natural transition orbitals we remove empty rows and columns from $\bm{T^{(2)}}$ then perform a singular value decomposition to give the natural transition geminals for the two states.

We construct the two-particle transition matrix by considering all SDs that comprise the two wavefunctions $\Psi_{2}=\sum_{i} c_{i} SD_{i}$ and $\Psi_{1}=\sum_{j} d_{j} SD_{j}$.  We note  that now $SD_{i}$ and $SD_{j}$ only contribute if they have two or fewer differences when in maximum coincidence.

If the two differences, in order of occurrence, in $SD_{i}$ are the spin orbitals $k_{1}$, $k_{2}$ while in $SD_{j}$ they are $l_{1}$ and $l_{2}$.  Then we have
\begin{equation}
T_{kl}\rightarrow T_{kl}+e_{g}e_{p}c_{i}^{*}d_{j}.
\end{equation}
Where $e_{p}$ is the sign from placing the SDs in maximum  coincidence and $e_{g}$ the sign from changing the order of $k_{1}$ and $k_{2}$ so that $k_{1} < k_{2}$ and similarly for $l_{1}$ and $l_{2}$ which are then
mapped to $k$ and $l$.

For one difference caused by spin orbital $k_{1}$ in $SD_{i}$ and $l_{1}$ in $SD_{j}$ then $k_{2}$=$l_{2}$ range over all the other spin orbitals in the Slater determinant and the order of, e.g., $k_{1}$ and $k_{2}$ is swapped if $k_{2}$ occurs before $k_{1}$ in the reordered Slater determinant.  This results in $N-1$ contributions which, when the labels are mapped to $k$ and $l$, are of the form
\begin{equation}
T_{kl}\rightarrow T_{kl}+e_{g}e_{p}c_{i}^{*}d_{j}.
\end{equation}

Finally we have the case of no differences. This results in

\begin{equation}
T_{mm}\rightarrow T_{mm}+e_{p}c_{i}^{*}d_{j}
\end{equation}
where $m$ ranges over all $\binom{N}{2}$ possible geminals which can be constructed from the spin orbitals of either of the Slater determinants.

\section{CH cation}

\begin{figure}
\begin{center}
\begin{minipage}{100mm}
\subfigure[]{
\resizebox*{5cm}{!}{\includegraphics{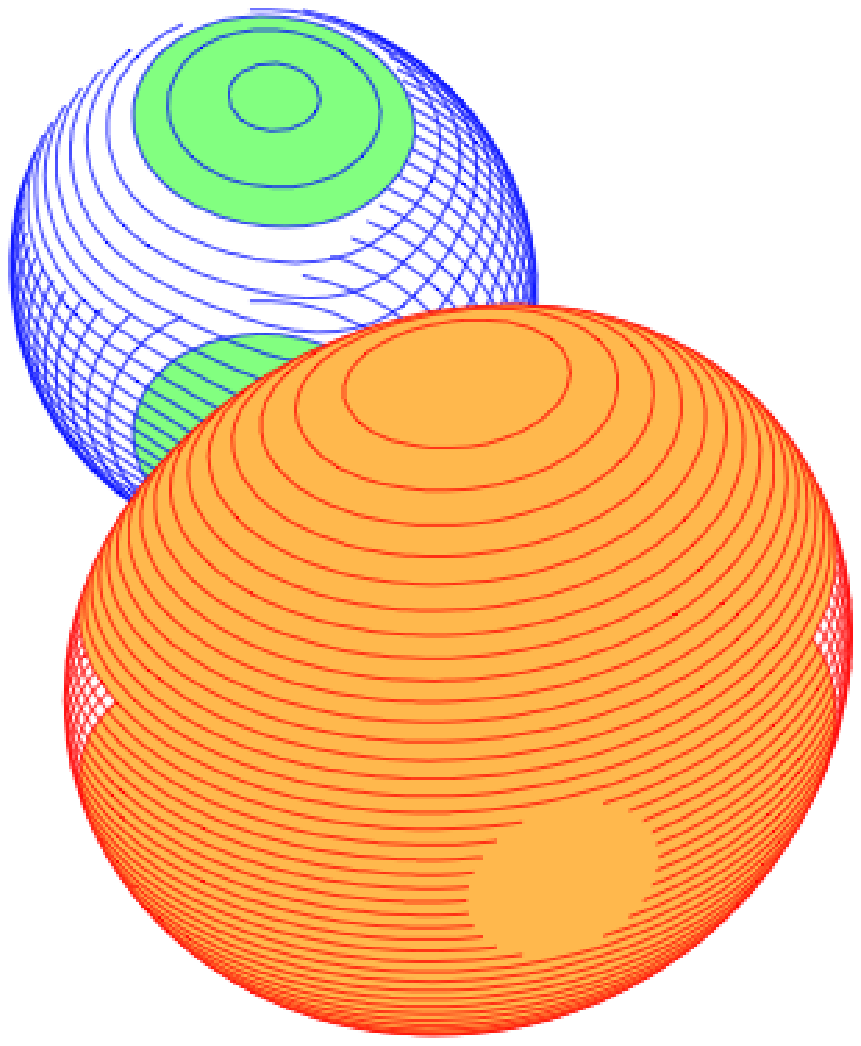}}}%
\subfigure[]{
\resizebox*{5cm}{!}{\includegraphics{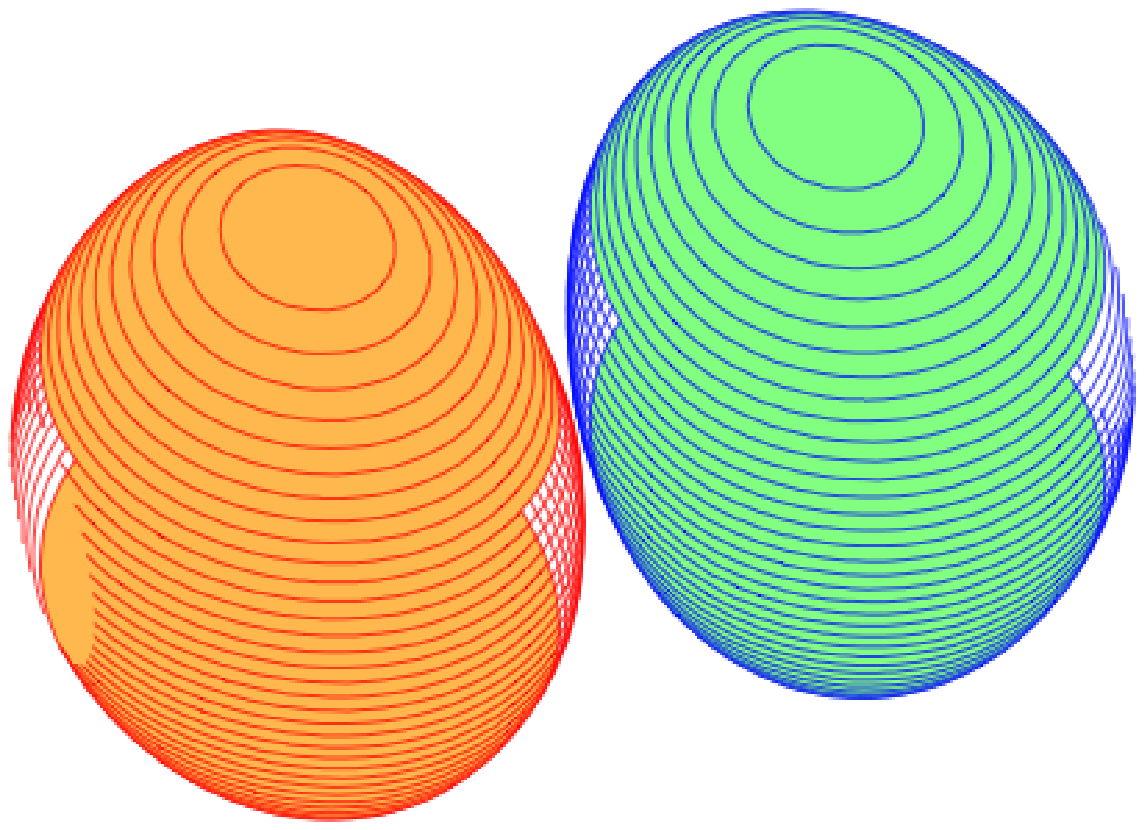}}}%
\caption{(a) HOMO $\phi_{3}$ of $A_{1}$ symmetry  (b) LUMOs are $\phi_{11}$ and $\phi_{15}$ of $B_{1}$ and $B_{2}$ symmetry respectively. Other LUMO is the same as that depicted but for a ninety degree clockwise rotation along the CH bond. }%
\label{fig:MOs}
\end{minipage}
\end{center}
\end{figure}

As an example of the use of natural transition geminals we now look at the ground and first singlet excited state of $A_{1}$ symmetry of the CH cation at a bond length of $1.13$ Angstrom. We use one frozen orbital for the CI calculations.  We analyze the CIS then the MCCI wavefunctions when using the cc-pVDZ basis.  The Hartree-Fock HOMO and LUMO, when plotted with Molden \cite{molden} using a 0.1 cutoff value, are displayed in Fig.~\ref{fig:MOs} where the carbon atom is at the front.

\subsection{Natural transition orbitals}

When using SDs, a CIS calculation gives the first excited state as a triplet here.  The second excited state has an energy of $-37.38$ Hartree which agrees with the Molpro \cite{Molpro} result
using configuration state functions (CSFs) of $-37.38$ so would be expected to be a singlet. For the natural transition orbitals we find that the main spin up and spin down transitions are the same with eigenvalue $0.71$ while the next transition has $3.8\times 10^{-2}$ for both spins and the other eigenvalues are negligible.  In the following results we neglect MOs with less than $10^{-1}$ contribution to the natural transition orbitals. The strongest transition is essentially $\phi_{3} \rightarrow 0.99\phi_{4}+0.12\phi_{7}$. This is then the HOMO to mainly the next orbital of $A_{1}$ symmetry.  The next transition is more complicated but is approximately $\phi_{2} \rightarrow 0.11\phi_{4} +0.64\phi_{5} -0.70\phi_{7} -0.25 \phi_{9} +0.18 \phi_{10}$.

The SA-MCCI calculation with $c_{\text{min}}=5\times 10^{-4}$ was run for 100 iterations on 4 processors.  This gave $-38.00$ and $-37.75$ Hartree for the ground and excited state respectively.  These energies agree to 2 d.p. with the FCI results using PSI3 \cite{PSI3} although the $1387$ SDS used by SA-MCCI represents a large fraction of the small FCI space of $6129$ SDs.

Now the three main spin up and spin down transitions have eigenvalues of $7\times 10^{-2}$ for the first two spin pairs and $4\times 10^{-2}$ for the third pair
while the others are less than $6\times 10^{-4}$.  Each up spin transition has the same relative phase as the down spin for each of these three confirming that the
excited state is not a triplet.  The first two transitions are 
\begin{eqnarray}
\nonumber 0.94\phi_{11}-0.20\phi_{12}-0.25\phi_{13}+0.16\phi_{14}\rightarrow -0.99\phi_{11}+0.15\phi_{12}, \\
\nonumber -0.94\phi_{15}+0.20\phi_{16}+0.25\phi_{17}-0.16\phi_{18}\rightarrow -0.99\phi_{15}+ 0.15\phi_{16},
\end{eqnarray}
while the third is $0.22\phi_{2}+ 0.97\phi_{3} \rightarrow \phi_{8}$.

With the MCCI wavefunction approaching the FCI here we see that excitations to the LUMO states are now some of the main transitions, but they are not from the HOMO as the symmetry must be preserved. Similarly one of the main transitions is from the HOMO but it is not to the LUMO states. However the natural transition orbitals only consider single excitations so we now compare these results with the natural transition geminals. 

\subsection{Natural transition geminals}

 In the following results we neglect geminals that contribute less than $10^{-1}$ to the natural transition geminals. When modelling the system using CIS and using natural transition geminals we find a transition with eigenvalue $0.999$: $\ket {3,\bar{3}}\rightarrow 0.70 \ket{3,\bar{4}}+0.70\ket{4,\bar{3}}$. 

There are then four with eigenvalue $0.71$:
 
\begin{eqnarray}
 \nonumber \ket {3,\bar{2}}\rightarrow 0.99 \ket{4,\bar{2}}+0.12\ket{7,\bar{2}}, \\ 
  \nonumber \ket{2,3}\rightarrow 0.99 \ket{2,4}+0.12\ket{2,7},\\
 \nonumber \ket{2,\bar{3}}\rightarrow 0.99 \ket{2,\bar{4}}+0.12\ket{2,\bar{7}}, \\
 \nonumber \ket{\bar{2},\bar{3}}\rightarrow 0.99 \ket{\bar{2},\bar{4}}+0.12\ket{\bar{2},\bar{7}}.
\end{eqnarray} 
Then there is one with eigenvalue $5.4\times 10^{-2}$:  $\ket{2,\bar{2}}\rightarrow 0.45 \ket{2,\bar{5}} -0.49 \ket{2,\bar{7}} -0.17 \ket{2,\bar{9}} +0.13 \ket{2,\bar{10}} + \left(\ket{i,\bar{j}} \leftrightarrow  \ket{j,\bar{i}} \right)$  while the rest are negligible.  We see that the $\phi_{3}$ to $\phi_{4}$ with some $\phi_{7}$ is the main single-particle transition similar to the results using natural transition orbitals.

For the SA-MCCI wavefunctions the main eigenvalue for the natural transition geminals is  $0.96$ while all the other eigenvalues are less than $9.63\times 10^{-2}$.  The dominant transition is now revealed as a double excitation which is mainly composed of the two HOMO electrons being excited to doubly occupied LUMOs
\begin{eqnarray}
\nonumber 0.13\ket{2,\bar{2}}-0.97\ket{3,\bar{3}}\rightarrow \\
\nonumber -0.69\ket{11,\bar{11}}+0.1\ket{11,\bar{12}}+0.1\ket{12,\bar{11}}+0.69\ket{15,\bar{15}}-0.1\ket{15,\bar{16}}-0.1\ket{16,\bar{15}}.
\end{eqnarray}

\section{Beryllium atom}

We now compare the SA-MCCI natural transition geminals with those of FCI for the electronic excitation between the first two  $A_{g}$ singlet states in the beryllium atom using a cc-pVDZ basis.  
We use no frozen orbitals and $D_{2h}$ symmetry.  A cutoff of $c_{\text{min}}=5\times10^{-3}$ is employed for the SA-MCCI calculation.  The Hartree-Fock orbitals have $\phi_{2}$ as the HOMO while the LUMOs are $\phi_{6}$, $\phi_{8}$ and $\phi_{11}$. The FCI result used 1093 Slater determinants to give an excitation energy of $7.75$eV.  We find that there is a transition with eigenvalue 1, four with 0.15, four with 0.12, four with 0.11 while the rest have eigenvalues lower than $0.02$. The largest eigenvalue corresponds to the transition
\begin{eqnarray}
\nonumber-0.95 \ket{2,\bar{2}} + 0.14 \ket{6,\bar{6}} +0.14 \ket{8,\bar{8}} +0.14 \ket{11,\bar{11}} \rightarrow 0.14 \ket{2,\bar{4}} +0.14 \ket{4,\bar{2}} \\ 
\nonumber  +0.58 \ket{6,\bar{6}} +0.18 \ket{6,\bar{7}}+0.18 \ket{7,\bar{6}} -0.67 \ket{11,\bar{11}} -0.21 \ket{11,\bar{12}} -0.21 \ket{12,\bar{11}}. 
\end{eqnarray}
For the SA-MCCI results we used 100 iterations and 4 processors. This gave an excitation energy of $7.76$eV using 27 Slater determinants. 
The eigenvalues for the natural transition geminals are similar to the FCI results: one transition with eigenvalue 1, four with 0.15, four with 0.13 and four with 0.10 while the rest have eigenvalues lower than $0.03$

The largest eigenvalue corresponds to the transition
\begin{eqnarray}
\nonumber -0.95 \ket{2,\bar{2}}+ 0.14\ket{6,\bar{6}} +0.14 \ket{8,\bar{8}} +0.14 \ket{11,\bar{11}} \rightarrow  \\
\nonumber 0.12\ket{2,\bar{4}}-0.10\ket{2,\bar{5}}+0.12\ket{4,\bar{2}}-0.10\ket{5,\bar{2}}+0.69\ket{6,\bar{6}}+0.22\ket{6,\bar{7}}\\
\nonumber+0.22\ket{7,\bar{6}}-0.14\ket{8,\bar{8}}-0.55\ket{11,\bar{11}}-0.17\ket{11,\bar{12}}-0.17\ket{12,\bar{11}}.
\end{eqnarray}

We see that when ignoring geminals with coefficient less than $0.1$ then the ground state natural transition geminals are the same for FCI and SA-MCCI with a reasonably large cutoff.  The excited state 
natural transition geminals have slightly different coefficients when using SA-MCCI and, notably, geminals with $\phi_{5}$ and $\phi_{8}$ are present in the SA-MCCI results but do not have a large enough coefficient in the FCI results to be displayed.  The transition cannot be straightforwardly described as involving a double excitation of MOs as the LUMOs also occur as doubly occupied geminals in the ground state albeit with small coefficients.  By considering the largest contributions however we can approximately describe the transition as comprising mainly a double excitation from the doubly occupied HOMO to the doubly occupied LUMOs of $B_{1u}$ and $B_{3u}$ symmetry. 

For the beryllium atom we also consider if the coefficients in the FCI expansion will reveal similar information about the important transition when using the natural orbitals of the ground state.  For the natural transition geminals we find that the eigenvalues are approximately the same  as when using the Hartree-Fock molecular orbitals.  The transition corresponding to the largest eigenvalue is
\begin{eqnarray}
\nonumber 0.95 \ket{2,\bar{2}}-0.17\ket{6,\bar{6}} -0.17\ket{8,\bar{8}}-0.17\ket{11,\bar{11}} \rightarrow  \\
\nonumber -0.11 \ket{2,\bar{4}} -0.11 \ket{2,\bar{5}} -0.11 \ket{4,\bar{2}} -0.11 \ket{5,\bar{2}} +0.18 \ket{6,\bar{6}} \\
\nonumber -0.74 \ket{8,\bar{8}}  +0.13 \ket{8,\bar{9}}+0.13 \ket{9,\bar{8}}+0.56 \ket{11,\bar{11}}.
\end{eqnarray}
We see that due to the use of the ground-state natural orbitals the coefficients are different particularly for some of the doubly occupied orbitals in the excited state.

Ordering the SDs in the FCI wavefunction by absolute coefficient gives for the ground-state
\begin{equation}
\nonumber \Psi_{0}=0.95\ket{1,\bar{1},2,\bar{2}}-0.17\ket{1,\bar{1},11,\bar{11}}-0.17\ket{1,\bar{1},8,\bar{8}}-0.17\ket{1,\bar{1},6,\bar{6}}+\dots  
\end{equation}

while for the excited we have

\begin{equation}
\nonumber \Psi_{1}=-0.74 \ket{1,\bar{1},8,\bar{8}}+0.56 \ket{1,\bar{1},11,\bar{11}}+0.18 \ket{1,\bar{1},6,\bar{6}}+\dots
\end{equation}

In this case one can deduce by inspection of the wavefunctions the main transition in the excitation and this is in agreement with the natural transition geminal result to two decimal places. Due to $\phi_{1}$ being doubly occupied in each of the main configurations then this becomes essentially a two particle system so it may be much more challenging to find by eye the constituents of the main transition in larger systems without recourse to the natural transition geminals.

\section{Carbon dimer}

 We now compare the natural transition geminals with the important configurations in a CI expansion for the carbon dimer.  Here we use a bond length of $2.348$ Bohr, the 6-31G* basis and a cut-off of $c_{\text{min}}=5\times10^{-3}$ for a SA-MCCI calculation.  The Hartree-Fock molecular orbitals are used as the system is sufficiently large that it is too computationally challenging to find the FCI natural orbitals.
 We freeze orbitals 1 and 15 hence they do not appear in any of the following expansions.  We calculate the ground and first excited state using 100 iterations on 4 processors.
The largest eigenvalue for the natural transition geminals is 0.92 while the rest are less than 0.12.  The corresponding transition is
\begin{eqnarray}
\nonumber 0.97 \ket{11,\bar{11}}-0.15\ket{8,\bar{8}}-0.13\ket{16,\bar{16}}-0.11\ket{25,\bar{25}} \rightarrow  \\
\nonumber 0.92 \ket{3,\bar{3}} -0.34 \ket{16,\bar{16}}. 
\end{eqnarray}

When sorting the CI expansions by the absolute value of the coefficient we find the following for the ground state

\begin{equation}
\nonumber \Psi_{0}=0.86 \ket{2,\bar{2},8,\bar{8},11,\bar{11},16,\bar{16}}-0.33\ket{2,\bar{2},3,\bar{3},8,\bar{8},11,\bar{11}}  +\dots  
\end{equation}

and for the excited state we see that
\begin{eqnarray}
\nonumber \Psi_{1}=0.92 \ket{2,\bar{2},3,\bar{3},8,\bar{8},16,\bar{16}}-0.17\ket{2,\bar{2},3,\bar{3},8,\bar{8},11,\bar{11}}\\
\nonumber -0.14\ket{2,\bar{2},3,\bar{3},11,\bar{11},16,\bar{16}}   +\dots  
\end{eqnarray}

Here it is not so straightforward to identify the dominant transition from the CI expansion and this would be expected to become more challenging if larger numbers of particles are considered.

\section{Formamide}

Finally we look at the $1$ $^{1}A' \rightarrow 2$ $^{1} A'$ transition of formamide. We use SA-MCCI running on 12 processors for  50 iterations with a cut-off of $c_{\text{min}}=5\times10^{-4}$.  The def1-TZVP basis \cite{def1TZVPbasisRef} is used and three core orbitals are frozen.  We use the MP2/6-31G* optimized geometry of Ref.~\cite{OrganicBenchmarks}.  The Hartree-Fock results give the HOMO as $\phi_{59}$, which is the second orbital of $A''$ symmetry, while the LUMO is $\phi_{11}$.  We find the excitation energy to the first
excited state to be $6.1$ eV here using $48832$ SDs compared with a FCI space of around $10^{21}$ SDs. In Ref.~\cite{OrganicBenchmarks} a CASPT2 result was $7.44$ eV while the CCSD response result appeared too low at $4.52$ eV. The natural transition geminals for the first excited state have one eigenvalue of 0.95 and forty-four of around 0.67 then the rest are less than $0.07$. The transition corresponding to the largest eigenvalue
is
\begin{equation}
\nonumber 0.99 \ket{59,\bar{59}} \rightarrow 0.69\ket{59,\bar{60}} - 0.69\ket{60,\bar{59}}.
\end{equation}
This suggests that the transition is a singlet to triplet so we look at the next excited state.  We now find an excitation energy of $8.4$ eV and the natural transition geminals now have one eigenvalue of 0.91, forty-four of $\sim0.6$, five around $0.1$ and the rest less than $0.1$. 
The largest eigenvalue is for the transition
\begin{equation}
\nonumber0.99 \ket{59,\bar{59}}\rightarrow 0.70\ket{59,\bar{60}} + 0.69\ket{60,\bar{59}}
\end{equation}
where we can see now that it is almost a singlet to singlet although there is a small amount of spin contamination.

The second largest eigenvalue of $0.67$ corresponds to the transition

\begin{eqnarray}
\nonumber  0.66\ket{59,\bar{58}} +0.75 \ket{58,\bar{59}} \rightarrow -0.22\ket{60,\bar{60}} \\
\nonumber +0.63 \ket{60,\bar{58}} -0.1 \ket{58,\bar{61}} -0.09  \ket{61,\bar{58}} +0.72 \ket{58,\bar{60}}.
\end{eqnarray}

Here we have included the $\ket{61,\bar{58}}$ geminal despite its coefficient being lower than 0.1 to show that the excitation is approximately singlet to singlet although this natural transition geminal seems to 
suggest slightly more spin contamination.  The large number of non-negligible eigenvalues here make it difficult to discuss what constitutes the main transition. However the results suggest that when modelled with SA-MCCI using SDs and a cut-off of $c_{\text{min}}=5\times10^{-4}$ then the transition to the first excited singlet
state of formamide of $A'$ symmetry with largest eigenvalue can be approximately thought of as a single excitation from the HOMO $\phi_{59}$ to $\phi_{60}$ although there is a small amount of double excitation character in the natural transition geminals of the second largest eigenvalue.

\section{Summary}

In this paper we introduced the use of natural transition geminals to aid in understanding excited states of configuration interaction wavefunctions that contain double excitations from
the ground-state.  We used the ground and first excited singlet of the CH cation as an initial example which we modelled using configuration interaction limited to single substitutions (CIS) and 
state-averaged Monte Carlo configuration interaction (SA-MCCI). The SA-MCCI results agreed with the full configuration interaction energies to two decimal places. Natural transition orbitals revealed that the excited state when using CIS could be essentially described as the HOMO ($\phi_{3}$) being 
excited to $\phi_{4}$ with some $\phi_{7}$.  Natural transition geminals were then seen to reproduce this information.  For the SA-MCCI wavefunctions, natural transition orbitals allowed the LUMOs ($\phi_{11}$ and $\phi_{15}$) to be viewed as some of the main orbitals created in the excited wavefunction during the transition but these came from the replacement of other orbitals of $B_{1}$ or $B_{2}$ symmetry.  This is due to the excitations being symmetry preserving so a double excitation would be needed to allow $A_{1}$ orbitals to be replaced with $B_{1}$.  Natural transition geminals then allowed the dominant transition for the SA-MCCI wavefunctions to be seen as a double excitation from mainly the doubly occupied HOMO to doubly occupied LUMOs.   

We saw that the FCI and SA-MCCI natural transition geminals were similar for the first singlet excitation of the beryllium atom when using the cc-pVDZ basis.  Here the SA-MCCI result used $27$ Slater determinants 
compared with $1093$ in the FCI calculation. Some geminals were not important enough to be in the FCI result yet were in the SA-MCCI result however the natural transition geminals showed that the transition was approximately dominated by a double excitation from the HOMO to two of the LUMOs in both cases even though the transition was somewhat complicated.  When using ground-state natural orbitals, we compared the transition suggested by the important configurations in the FCI expansions with the natural transition geminals and found for this system that they agreed. We then considered the carbon dimer and saw that it was more difficult to 
identify the important transition by looking at the CI expansion for this larger system.

The natural transition geminals allowed us to identify the first excited state of formamide as a triplet when using SA-MCCI with Slater determinants.  The second excited state of $A'$ symmetry appeared to be
approximately a singlet, although some spin contamination was observed in the natural transition geminals, and there was a large number of significant eigenvalues suggesting that there may not be a simple interpretation of the transition. The two largest eigenvalues had a single particle
excitation from the HOMO to another orbital of the same symmetry as the main contribution here although the transition for the second largest eigenvalue had some double excitation character.

Due to the number of geminals scaling as $O(M^{2})$, where $M$ is the number of orbitals, then calculating the natural transition geminals by performing a singular value decomposition can become onerous
as the size of the basis becomes large, although we note that natural transition geminals could be used for most CASSCF wavefunctions due to the restriction on the number of orbitals in the active space.  Furthermore, as mentioned in the introduction, the use of geminals is more complicated than orbitals and we do not know of a useful approach to visualize the natural transition geminals.  Despite these difficulties, it appears that the natural transition geminals can be useful for giving a qualitative understanding of the 
main transitions to excited states where single and double excitations are important when using configuration interaction wavefunctions.

\section*{Acknowledgement} We thank the European Research Council (ERC) for funding under the European Union's Seventh Framework Programme (FP7/2007-2013)/ERC Grant No. 258990.

\providecommand{\noopsort}[1]{}\providecommand{\singleletter}[1]{#1}%

\end{document}